# Condition Sensing for Electricity Infrastructure in Disasters by Mining Public Topics from Social Media

**Yudi Chen**
George Mason University
ychen55@ gmu.edu

**Angel Umana**
George Mason University
aumana2@gmu.edu

**Chaowei Yang**
George Mason University
cyang3@gmu.edu

**Wenying Ji**
George Mason University
wji2@gmu.edu

**ABSTRACT**

Timely and reliable sensing of infrastructure conditions is critical in disaster management for planning effective infrastructure restorations. Social media, a near real-time information source, has been widely used in disasters for forming timely situational awareness. But using social media to sense electricity infrastructure conditions has not been explored. This study aims to address the research gap through mining public topics from social media. To achieve this purpose, we proposed a systematic and customized approach wherein (1) electricity-related social media posts are extracted by the classifier developed from the Bidirectional Encoder Representations from Transformers (BERT); and (2) public topics are modeled with unigrams, bigrams, and trigrams to incorporate the formulaic expressions of infrastructure conditions in social media. Electricity infrastructures in Florida impacted by Hurricane Irma are studied for illustration and demonstration. Results show that the proposed approach is capable of sensing the temporal evolutions and geographic differences of electricity infrastructure conditions.

**INTRODUCTION**

Electricity infrastructure is the fundamental underpinning of modern societies as it supports all other critical infrastructures (e.g., water treatment, telecommunications, and public health) in fulfilling the basic needs of the public (DOE, 2017). Recently, natural disasters (e.g., earthquakes, hurricanes, and tornadoes) have occurred with an increasing frequency and intensity due to climate change, and subsequently brought massive electricity disruptions (Bartos & Chester, 2015). The massive electricity disruptions are further aggravated by the aging electricity infrastructure in the United States (US); it is graded as level D (representing poor to fair conditions) as per the 2017 American Society of Civil Engineers (ASCE) infrastructure report card (ASCE, 2017). To achieve effective restorations, timely and reliable sensing of electricity infrastructure conditions is needed for practitioners to make informed decisions (e.g., crew dispatch and information update).

As the aggregated information of human opinions, public topics are promising for providing useful information related to electricity infrastructure conditions. Recently, social media has been widely used in disaster management for mining public topics, which contributes to the development of rapid and reliable awareness of general disaster situations (DHS, 2014; Rexiline Ragini, Rubesh Anand, & Bhaskar, 2018). Public topics in different stages of Hurricane Irma are identified from Twitter to understand general disaster situations (e.g., caution and advice, the safety of people and animal, and infrastructure status), which is beneficial for emergency managers to timely and effectively address general public needs (Xu, Lachlan, Ellis, & Rainear, 2019). Highway-related public topics are identified from Twitter to obtain a rapid and reliable assessment of disaster impacts on highways from Hurricane Harvey, which contributes to effective highway recovery and route planning (Chen, Wang, & Ji, 2020). Infrastructure-related topics are detected from Twitter to track the evolutions of disaster situations, which is beneficial for providing actional insights to responders (Fan, Mostafavi, Gupta, & Zhang, 2018). Public health topics are mined from various social media platforms (e.g., Twitter, Facebook, and Instagram) to support the preparedness and response of public health infrastructure in the currently evolving pandemic COVID-19, which is useful to prevent hospitals from being overwhelmed (Merchant & Lurie, 2020). Such success of presented outcomes demonstrates the feasibility and utility of social media-based public topics in providing or enhancing situational awareness in disasters. Despite being an intriguing idea, using social media-based public topics for sensing electricity conditions has yet to be explored mainly due to the lack of a specific social media mining approach.

This research aims to address this research gap by proposing a systematic and customized approach to sense electricity infrastructure conditions through mining public topics from social media. Explicitly, the developed approach is capable of (1) accurately extracting electricity-related social media data with binary classifier developed based on Bidirectional Encoder Representations from Transformers (BERT); (2) reliably modeling public topics on electricity infrastructure with the most frequent unigrams, bigrams, and trigrams to incorporate the formulaic expressions used for describing electricity conditions in social media; and (3) systematically investigating the spatiotemporal patterns of the modeled public topics to indicate electricity infrastructure conditions. Electricity infrastructures in Florida impacted by Hurricane Irma are studied for illustration. The content of this paper is organized as follows. First, research studies focusing on text representation techniques are reviewed. Second, the details of the developed approach are introduced. Third, the spatiotemporal patterns of the modeled public topics are analyzed. Finally, research contributions, limitations, and future work are discussed.

**TEXT REPRESENTATION**

As the main information format, social media text has been extensively used in disaster management to derive timely and reliable disaster situations (Alexander, 2014; Caragea, Silvescu, & Tapia, 2016; Imran, Castillo, Diaz, & Vieweg, 2018; Yin et al., 2015). To process the text information efficiently, text mining techniques are needed to extract the information related to a specific topic (i.e., electricity infrastructure in this research) from massive social media data.

Text representation is a critical and fundamental problem in text mining techniques (Aggarwal & Zhai, 2012). Unstructured texts should be represented with numerical values to make them mathematically computable. Bag of words (BOW) and term frequency-inverse document frequency (TFIDF) are two traditionally used text representation methods. In BOW, a text is represented as the bag of its words, and the value for each word is its frequency (Rajaraman & Ullman, 2011). TFIDF has a similar format to BOW, but its word frequency is normalized by the document frequency (Rajaraman & Ullman, 2011). However, these two traditional methods are incapable of capturing semantic meanings, as they disregard the valuable word order information. Recently, BERT has been increasingly used to build an accurate text representation due to its superiority of semantic meaning interpretation (Devlin, Chang, Lee, & Toutanova, 2018). BERT is designed to pre-train deep bidirectional representations from unlabeled texts by jointly conditioning on both left and right contexts in all layers. The pre-trained BERT model can be fine-tuned with just one additional output layer to adapt to the application corpus. BERT has been used to create state-of-the-art models for a wide range of tasks (e.g., question answering and text classification) without substantial task-specific architecture modifications (Devlin et al., 2018). Several existing studies have demonstrated the capability of BERT in disaster context for building accurate classifiers on classifying social media data into various humanitarian categories (e.g., impacted individuals, infrastructure damage, and other relevant) (Fan, Wu, & Mostafavi, 2020) and identifying flood events (de Bruijn et al., 2019). Such presented research outcomes, as well as the superiorities of BERT in text representation, make it promising to build an accurate BERT-based classifier to identify electricity-related tweets in disasters.

**METHODOLOGY**

The developed approach consists of four modules: social media data collection, electricity-related data classification, electricity-related public topic modeling, and electricity infrastructure condition sensing. In the social media data collection module, Twitter is employed to collect raw social media data due to its popularity in disaster management, as well as its easy access for collecting a large-scale dataset (Kryvasheyeu et al., 2016). The raw collected data is further filtered with a set of carefully designed electricity-specific keywords to reduce information overload. The purpose of reducing information overload is to ease the manual efforts required for labeling a high-quality training dataset in the following classification module. In the electricity-related data classification module, the identification of electricity-related social media data is formed as a binary classification problem: each of the collected tweets is classified as electricity-related or non-electricity-related. Using the identified electricity-related social media data, public topics on electricity infrastructure are modeled by the most frequent unigrams (single word), bigrams (continuous sequence of two words), and trigrams (continuous sequence of three words) to incorporate the formulaic expressions used by social media users for describing their electricity conditions in emergency environments. Finally, the sensing of electricity infrastructure conditions is conducted by investigating the temporal and geographic distributions of the modeled public topics.

Public topics on the electricity infrastructure in Florida impacted by Hurricane Irma are studied for illustration. Hurricane Irma was an extremely powerful hurricane that caused widespread destruction across its path in September 2017. 6.7 million electricity customers, which were 64% of all customer accounts in Florida, were with power outages (EIA, 2018). Hurricane Irma became a Major Hurricane (Category 3) on September $1^{st}$, 2017, made its first landfall in the Florida Keys on September $10^{th}$, 2017, traveled up to Florida on September $11^{th}$, 2017, and finally degenerated to a remnant low and moved away from Florida on September, $12^{th}$, 2017 (Cangialosi, Latto, & Berg, 2018). According to the timeline of Hurricane Irma, we defined three phases to capture the temporal evolutions of Hurricane Irma: before-Irma, during-Irma, and after-Irma. Explicitly, the before-Irma phase is from September $1^{st}$, 2017 to September $9^{th}$, 2017, the during-Irma phase is from September $10^{th}$, 2017 to September $11^{th}$, 2017, and the after-Irma phase is from September $12^{th}$, 2017 to September $30^{th}$, 2017.

**Social Media Data Collection**

Tweepy, a Python package for implementing the Twitter streaming API (Roesslein, 2020), is used to collect geotagged tweets through the previously developed module (Q. Wang & Taylor, 2016). Two types of constraints, timespan and location, are used as filters to ensure that the collected tweets were posted during the period of Hurricane Irma and in disaster-impacted regions. The timespan is from September $1^{st}$, 2017 to September $30^{th}$, 2017 to cover the predefined three disaster phases. The location is set as the state of Florida as it is the most severely impacted state. Finally, 348,629 tweets posted by 43,644 unique users are collected.

The collected tweets are further filtered with a set of electricity-specific keywords to reduce information overload. Previous studies have proved that the information related to a specific infrastructure in social media only accounts for a small portion, and keyword filtering is an effective way of removing most of the irrelevant social media data (Chen, Ji, & Wang, 2020). To ensure the completeness of electricity-related information, the employed keywords should be carefully designed to cover the most mention scenarios of electricity infrastructure in social media. Through manually checking the collected tweets, seven keywords (blackout, electric, electricity, outage, power, dukeenergy, and FPL) are employed to cover both the general and local mention scenarios of electricity infrastructure. Keywords of "blackout", "electric", "electricity", "outage", and "power" are used to cover general mention scenarios, as they have no specific location information. Keywords of "dukeenergy" and "FPL", which are the names of two main utility companies (Duke Energy and Florida Power & Light) in Florida, are employed to cover the local mention scenarios. Notably, the abbreviations of these keywords, such as "pwr" to "power", are also included as the keywords. Additionally, to remove the tweets posted by agencies or even bots, the tweets posted by extremely active Twitter users are removed through empirically setting a threshold, i.e., tweeting frequency is greater than 10 in the studied period. Through this electricity-specific keyword filtering, 2,291 tweets (0.66% of the initially collected tweets) are finally kept, which signifies a considerable reduction of information overload. The number of filtered tweets for each keyword is shown in Figure 1. To allow a large range, the x-axis is on a log scale with a base of 10. The tweets filtered by the keyword "power" are the majority. Notably, the filtered tweets are not necessarily related to electricity infrastructure. For example, although the tweet "*In our weaknesses, God's power and strength are truly manifested*" contains "power" that is a keyword for electricity infrastructure, it is apparently not related to electricity infrastructure. Therefore, further identification of electricity-related tweets from the filtered tweets is still needed.

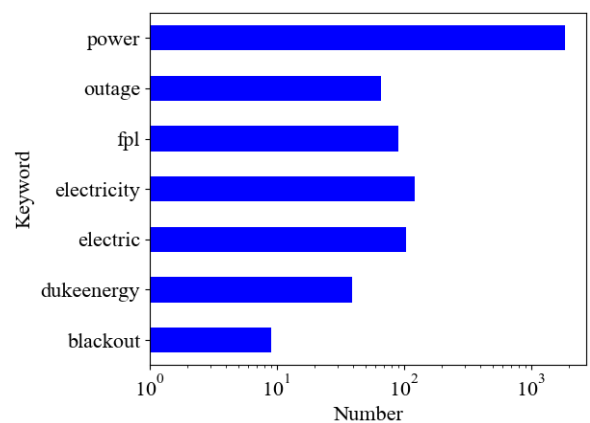

**Figure 1. Number of the Filtered Tweets**

**Electricity-related Data Classification**

In this section, the identification of electricity-related tweets is formed as a binary classification problem. A fine-tuned BERT-based classification model is developed to classify the filtered tweets as electricity-related or non-electricity-related. The architecture of the BERT-based binary classification model is depicted in Figure 2. An input tweet is tokenized into *N* tokens: Tok 1 to Tok N. [CLS] is a special symbol added in front of every input text. Ts are the final output of the input text token. C is the final output of the special symbol [CLS], and it is usually used as the aggregate sequence representation for classification tasks. In addition to BERT, an additional output layer is added on top of the output C to predict class labels. Since the electricity-related tweet classification is binary, the output layer is only a sigmoid neuron (S). The sigmoid function calculates the probability of a class given an input vector $x$, and it is formulated as Equation (1).

$$S(x) = \frac{1}{1 + e^{-wx^{T}}} \quad (1)$$

where $w$ is the weights (the parameters in the output layer). The BERT-based classification model is simply composed of BERT and only an additional sigmoid neuron, which significantly reduces the efforts required for building deep learning architectures. In this research, the pre-trained BERT model with 12 layers (https://tfhub.dev/tensorflow/bert_en_uncased_L-12_H-768_A-12/1) is used for initialization. During the training process, the pre-trained parameters in BERT are fine-tuned, and the parameters in the output layer (connects the output C and the sigmoid neuron S) are trained.

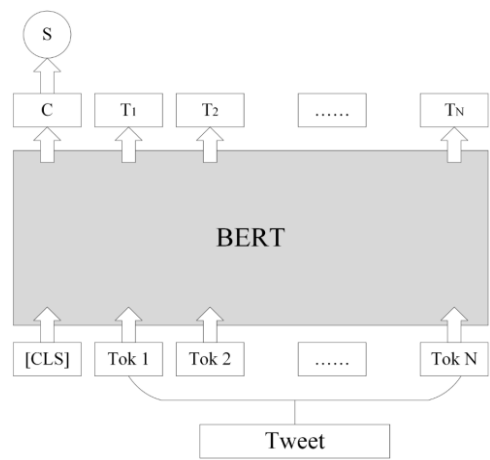

**Figure 2. Architecture of the BERT-based Binary Classification Model**

To illustrate the superiority of the BERT-based model for this binary classification task, support vector machine (SVM) (Suykens & Vandewalle, 1999) and logistic regression (LR) (Kleinbaum & Klein, 2002), which are two well-established text classification approaches, are included as baselines. BOW and TFIDF are used to transform social media text into numerical features as the input of SVM and LR. In this research, an electricity-related tweet is taken as a positive sample as this research focuses on electricity infrastructure, and consequently, a non-electricity-related tweet is taken as a negative sample. Precision, recall, and $F_1$ are employed to evaluate performances. A relevant sample represents an electricity-related tweet.

A training dataset, wherein each tweet is labeled as electricity-related or non-electricity-related, is needed to train these supervised classifiers. The reduction of information overload in the above section provides a valuable opportunity to obtain a high-quality training dataset (i.e., rich electricity-related information) through labeling only a small number of tweets, which significantly reduces the required manual labeling efforts. The reduction of information overload guarantees that a large portion of the filtered tweets is related to electricity infrastructure, as they contain at least one of the electricity-specific keywords. To ensure labeling accuracy, an English native speaker was invited to manually label 1,000 tweets, wherein 762 tweets are electricity-related, and the remaining 238 tweets are non-electricity-related. The labeled 1,000 tweets are randomly split into three subsets: training (60%), validation (20%), and testing (20%) subset. The training and validation subsets are used to tune the hyperparameters of supervised classification models, such as the learning rate in the BERT-based classification model, and the penalties in SVR and LR. Once the best hyperparameters for a model are found, the model with the best hyperparameters is retrained with the combination of training and validation subsets, then tested on the testing subset. Google Colab, which provides free access to Google's hardware (GPUs and TPUs), is employed in this research for training and testing (Bisong, 2019). Table 1 shows the testing performances of the BERT-based model and baseline models, and the best performance is marked in bold. The BERT-based model achieves the best performances: all evaluation performance metrics are around 97%. The BERT-based classifier outperforms the baselines, especially in terms of precision. Through manual checking of the labeled tweets, we found that lots of formulaic expressions are used by social media users to describe their electricity conditions, such as "have power", "lose power", and "no power". This phenomenon makes BERT an ideal candidate for this binary classification task, as it is capable of learning these formulaic expressions by considering bidirectional word sequences and fine-tuning its parameters to adapt to this application.

**Table 1. Comparisons of classification performance**

| Model | Precision | Recall | $F_1$ |
|---|---|---|---|
| LR-BOW | 87.8% | 96.4% | 91.9% |
| LR-TFIDF | 91.5% | 98.0% | 94.7% |
| SVM-BOW | 89.9% | 98.0% | 93.9% |
| SVM-TFIDF | 87.4% | 98.0 % | 92.5% |
| **BERT-based** | **96.8%** | **98.1%** | **97.4%** |

With the fine-tuned BERT-based classification model, each of the remaining 1,291 tweets (2,291 filtered tweets – 1,000 labeled tweets) is classified as electricity-related or non-electricity-related, of which 1,001 tweets are classified as electricity-related. The combination of the electricity-related tweets obtained through the fine-tuned BERT-based model and the electricity-related tweets in the labeling dataset are combined as the final dataset related to electricity infrastructure. Finally, 1,763 (i.e., 1,001 + 762) electricity-related tweets are extracted, and their temporal variations are shown in Figure 3. Electricity-related Twitter activities remained at a low level before the landfall of Irma, increased sharply during Irma, and gradually decreased in the after-Irma phase.

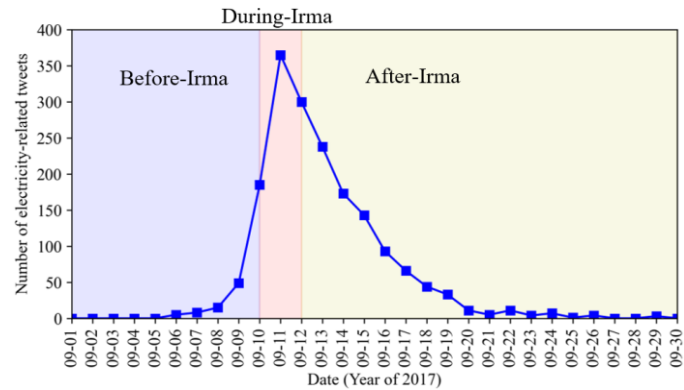

**Figure 3. Temporal Variations of Electricity-related Twitter Activities**

**Electricity-related Public Topic Modeling**

In this research, electricity-related public topics are modeled with the most frequent terms, and a term refers to a unigram, bigram, or trigram. Previous studies have demonstrated the utilities of bigrams and trigrams in language models for speech recognition (Mikolov, Sutskever, Chen, Corrado, & Dean, 2013; X. Wang, McCallum, & Wei, 2007). Meanwhile, in the electricity-related data classification module, we found that lots of bigrams and trigrams are particularly used as formulaic expressions by social media users to describe their electricity conditions, such as "have power" and "still no power". Therefore, bigrams and trigrams are incorporated together with unigrams to model electricity-related public topics.

To achieve reliable modeling of electricity-related public topics, a systematic process for identifying the most frequent *k* terms (hereafter refers to top-*k* terms for simplicity) is defined, as shown in Figure 4. In the tweet cleaning and tokenization, all URL links and invalid symbols are removed, and then each tweet is tokenized into a set of separate tokens. The tokens are lemmatized into their base forms, such as "powered" to "power". The bigrams and trigrams are built based on the entire electricity-related tweets with the constraints of minimum count and scoring threshold. The minimum count constraint is used to ignore the bigrams and trigrams with a small total collected count: the minimum count is empirically set as 5. The scoring threshold is used to ignore the bigrams and trigrams that are unlikely to be phrases. For a bigram composed of the word a (the first word) and b (the second word), its score is calculated with Equation (2) (Mikolov et al., 2013).

$$\text{score}(a,b) = \frac{(C_{bigram} - C_{\min}) * n}{C_a * C_b} \quad (2)$$

The $C_{bigram}$, $C_a$, and $C_b$ are the counts of the bigram, word a, and b, respectively. $C_{\min}$ is the value of the minimum count constraint, and it is used as a discounting coefficient to prevent too many phrases consisting of very infrequent words to be formed. *n* is the size of the corpus built using the entire electricity-related tweets. The trigrams are constructed based on the built bigrams and remaining unigrams, i.e., the combination of a built bigram and a unigram is a trigram. A higher score threshold means fewer phrases. In this research, the score threshold is empirically set as 1 to balance the numbers and the meanings of bigrams and trigrams. Once bigrams and trigrams are built, the stop words are removed, as stop words alone provide no meanings. The stop words are from the Natural Language Toolkit (NLTK, 2007). Here, the useless tokens are the stop words (meaningless information) and the previously employed electricity-related keywords (already known information) that remain as unigrams. FPL and Dukeenergy are merged into a single bigram, i.e., utility company, as they have no differences in understanding public topics on electricity infrastructure.

```
Input: Electricity-related tweets
1. Clean and tokenize electricity-related tweets
2. Lemmatize the tokens
3. Build bigrams and trigrams
4. Remove useless tokens
5. For each disaster phase
6.    Load the tweets posted on the disaster phase
7.    Identify the top-k terms
8. End
Output: Top-k terms for each disaster phase
```

**Figure 4. Top-k Terms Identification Process**

**Electricity Infrastructure Condition Sensing**

In this section, the modeled public topics are investigated from both temporal and spatial perspectives. The temporal perspective is used to indicate the evolution of electricity infrastructure conditions, which contributes to the interpretation of electricity conditions in different disaster phases, i.e., before-, during-, and after-disaster phases. Additionally, such interpretation enables the understanding of the interplay between human behaviors and infrastructure conditions. The spatial perspective of the modeled public topics is used to compare electricity infrastructure conditions amongst different geographic regions, such as the impacted counties in a state. For example, by comparing the intensity of electricity infrastructure damage-related public topics, we could identify the most severely impacted regions as they usually have a high intensity of damage-related mentions on social media. By doing such spatial comparisons, we are capable of perceiving the geographic region that has the most electricity infrastructure disruptions, as well as the regions wherein the electricity infrastructures are only slightly impacted. The details of utilizing the modeled public topics to sense electricity infrastructure conditions are illustrated in the results.

**RESULTS**

In this section, the utilization of the modeled public topics for sensing electricity infrastructure conditions are conducted through a two-step procedure: (1) the top terms which are with clear indications of electricity conditions are grouped into five aggregated topics (i.e., no-power, have-power, safety-check, damage, and restoration), then (2) the spatiotemporal patterns of the aggregated topics are studied over the counties with significant electricity-related Twitter activities. Through such a procedure, spatiotemporal patterns of public topics are examined from an aggregate level, which is needed for achieving reliable sensing of electricity infrastructure conditions.

**Step 1: Top Term Aggregation**

In this research, the *k* is set as 20 to balance the number and the topic coverage of top terms. The top 20 terms in the three predefined disaster phases are listed in Table 2. The number after a disaster phase is the number of tweets in this phase, and the number after a top term is the count of this top term. In the before-Irma phase, the term "power outage", "lose power", and "power out" emerged, which indicated that there were electricity disruptions in certain communities even before the landfall of Hurricane Irma. The term "ready" and "cook" reflected the public preparation for Hurricane Irma, and cooking food is the most popular preparation behavior. In the during-Irma phase, the terms related to electricity disruptions, such as "no power" and "lose power", were quite popular (with large numbers), which indicated that Hurricane Irma brought massive electricity disruptions. Meanwhile, "still have power" was the second most popular term, and this indicated that certain communities survived Hurricane Irma without losing electricity. The term "good" and "safe" reflected safety check behaviors, and the term "tree" and "damage" described the damage caused by Hurricane Irma on the electricity infrastructure of Florida. In the after-Irma phase, "have power" became the most popular term, as power came back in most communities. This was partially demonstrated by the high popularity of the term "power back". Apart from the terms related to electricity disruptions used in the before- and during-Irma phases, "still no power" emerged in the after-Irma phase. The term "still no power" revealed that certain communities were without power for a long period, which indicated that the communities were severely impacted by Hurricane Irma and long restoration processes were needed. The electricity conditions indicated by the term "still no power" is exactly opposite to that indicated by the term "still have power". Additionally, the terms related to restoration processes began to emerge, such as "utility company" and "restore", in the after-Irma phase.

To better understand public topics on electricity infrastructure, only the top terms that are with obvious indications of electricity conditions are considered for further analysis. Five aggregated topics are empirically prepared, as shown in Table 3. The first two aggregated topics are no-power and have-power which indicate two opposite electricity conditions. Understanding the public topics of the two sides of electricity conditions is capable of assisting responders to develop a more comprehensive and reliable situational awareness. The remaining three aggregated topics are safety-check, damage, and restoration. Safety-check is a popular public behavior in disasters for checking the safety of their friends and family, and it is useful for understanding public situations under massive electricity disruptions. The information related to electricity infrastructure damage is beneficial for responders to perform damage assessment that is needed for dispatching utility crews effectively. The restoration topic can be potentially used to evaluate electricity recovery process from a human-centered perspective.

**Table 2. Top-*k* Terms in the Three Disaster Phases**

| Rank | Before-Irma (77) | During-Irma (550) | After-Irma (1136) |
|---|---|---|---|
| 1 | ('power outage', 11) | ('no power', 68) | ('have power', 101) |
| 2 | ('lose power', 9) | ('still have power', 52) | ('no power', 86) |
| 3 | ('ready', 7) | ('still', 48) | ('home', 67) |
| 4 | ('nb sb', 7) | ('lose power', 48) | ('still no power', 62) |
| 5 | ('power out', 6) | ('tree', 43) | ('day', 62) |
| 6 | ('go', 6) | ('good', 40) | ('utility company', 61) |
| 7 | ('cook', 6) | ('safe', 39) | ('without power', 60) |
| 8 | ('sfltraffic', 6) | ('power out', 37) | ('power back', 59) |
| 9 | ('still have power', 5) | ('wind', 36) | ('power outage', 58) |
| 10 | ('the avenue', 5) | ('go', 31) | ('back', 58) |
| 11 | ('street', 5) | ('damage', 31) | ('still', 51) |
| 12 | ('power go out', 5) | ('power go out', 28) | ('restore', 45) |
| 13 | ('good', 4) | ('have power', 28) | ('work', 45) |
| 14 | ('wind', 4) | ('power outage', 26) | ('open', 45) |
| 15 | ('still', 4) | ('without power', 25) | ('water', 42) |
| 16 | ('make', 4) | ('make', 22) | ('get power', 42) |
| 17 | ('stay', 4) | ('power line', 21) | ('today', 39) |
| 18 | ('forecast', 4) | ('road', 20) | ('thank you', 38) |
| 19 | ('home', 3) | ('house', 20) | ('week', 35) |
| 20 | ('tomorrow', 3) | ('time', 20) | ('house', 35) |

**Table 3. Five Aggregated Topics**

| Aggregated Topic | Included top terms |
|---|---|
| No-power | 'power outage', 'lose power', 'no power', 'without power', 'still without power', 'power out', 'still no power', 'power go out' |
| Have-power | 'have power', 'still have power', 'power back on', 'power back', 'get power' |
| Safety-check | 'good', 'safe' |
| Damage | 'tree', 'damage', 'fall power cable', 'power line' |
| Restoration | 'utility company', 'restore' |

The count of an aggregated topic is the sum of the counts of its included terms. The temporal variations of the five aggregated topics are shown in Figure 5. Topic engagement *TE* is defined to reveal public participation in an aggregated topic, and it is formulated as Equation (3).

$$TE_{T,P} = \frac{ATC_{T,P}}{NT_P} \qquad (3)$$

where *ATC* and *NT* are the count of an aggregated topic and the number of tweets, respectively. The subscripts *T* and *P* represent the aggregated topic and the disaster phase, respectively. The no-power aggregated topic was the most popular in all three disaster phases, and the have-power aggregated topic followed. The *TE* of the no-power aggregated topic had similar values in the before-Irma and during-Irma phases but decreased significantly in the after-Irma phase. This indicated that high percentages of electricity-related tweets in the before- and during-Irma phases described electricity disruptions. In contrast to the temporal variation of the no-power topic, the have-power aggregated topic increased significantly from the before-Irma phase to the during-Irma phase and remained at a similar level in the after-Irma phase. This revealed that have-power topic was popular not only in the after-Irma phase but also in the during-Irma phase: the popularity in the after-Irma phase was caused by the electricity restoration, while the popularity in the during-Irma phase was caused by the situation update posted by the users who survived Hurricane Irma without losing electricity. The safety-check and damage topics achieved their peaks in the during-Irma phase, which reflected that the public usually did the safety-check and shared the damage on the first time. The restoration topic peaked in the after-Irma phase, as utility crews were dispatched for restoring electricity infrastructure.

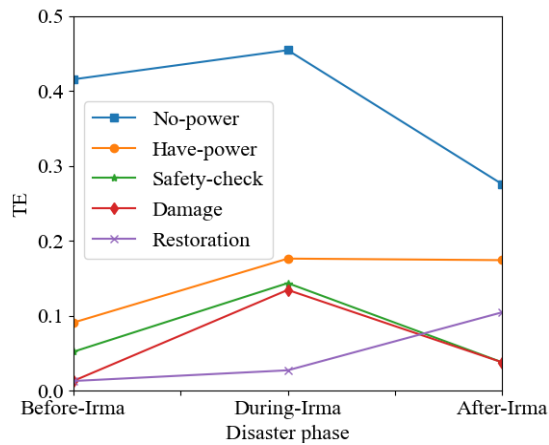

**Figure 5. Temporal variations of the five aggregated topics**

**Step 2: Spatiotemporal Patterns**

To ensure a reliable analysis, only counties with significant electricity-related Twitter activities are employed to study their spatiotemporal patterns. The geographic distributions of electricity-related Twitter activities in Florida are shown in Figure 6. Notably, this figure is only used to provide an overview of the geographical distribution of electricity-related tweets, and disaster impact assessment is presented in Figure 7. The strength of Twitter activities is represented by the number of electricity-related tweets. Most counties are with very few electricity-related tweets, even on the path of hurricane Irma. Among the 67 counties in Florida, 45 counties have very few (<10) electricity-related tweets. Counties possessing a large number of electricity-related tweets (>=100) include Miami-Dade (287), Broward (215), Orange (186), Pinellas (140), and Palm Beach (110). These five counties are critical counties of three metropolitan areas in Florida. Miami-Dade, Broward, and Palm Beach County are the three counties in the Miami metropolitan area. Orange County is the central county of the Orlando-Kissimmee-Sanford metropolitan area. Pinellas County is a part of the Tampa-St. Petersburg-Clearwater metropolitan area. Considering that cities performed as public activity centers, electricity-related Twitter activities gathering in these counties, in this case, is reasonable. These five counties are further employed to study the spatiotemporal patterns of the five aggregated topics.

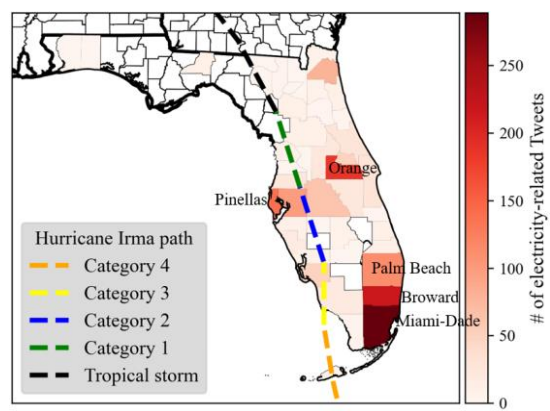

**Figure 6. Geographic Distribution of Electricity-related Twitter Activities**

Spatiotemporal analysis of urban settings has been frequently used in disaster management to understand disaster dynamics (Yu et al., 2020). In this research, the spatiotemporal patterns of the five aggregated topics over the five counties in the three predefined disaster phases are illustrated with a group bar plot, as shown in Figure 7. Each county is represented by a color. The topic engagement TE in a county is calculated based on the electricity-related tweets only posted in this county. In the before-Irma phase, no-power aggregated topic in Miami-Dade, Broward, and Palm Beach Counties dominated, and it is reasonable as these three counties were firstly impacted by Hurricane Irma. There were no electricity-related Twitter activities related to the topic of damage and restoration. In the during-Irma phase, no-power aggregated topic was still dominant in all counties. In the no-power topic, Pinellas County had the highest TE, while in the have-power topic, Pinellas County had the lowest TE. This is consistent with the fact that Pinellas County was severely impacted by Hurricane Irma, and was with massive electricity disruptions (Johnston & Griffin, 2017). In the have-power topic, Orange County had the highest TE, and this is because the percentage of customers without power in Orange County was relatively lower than that in the other four counties (Johnston & Griffin, 2017). Pinellas County had the highest TE in the safety-check aggregated topic. The differences in the damage aggregated topic were not significant. In the after-Irma phase, there were not many differences in the five aggregated topics, except for the restoration topic. In the restoration topic, the TE of Pinellas County was way higher than the TEs in the other four counties. This can be explained by the fact that Pinellas County was severely impacted by Hurricane Irma, and therefore required more restoration.

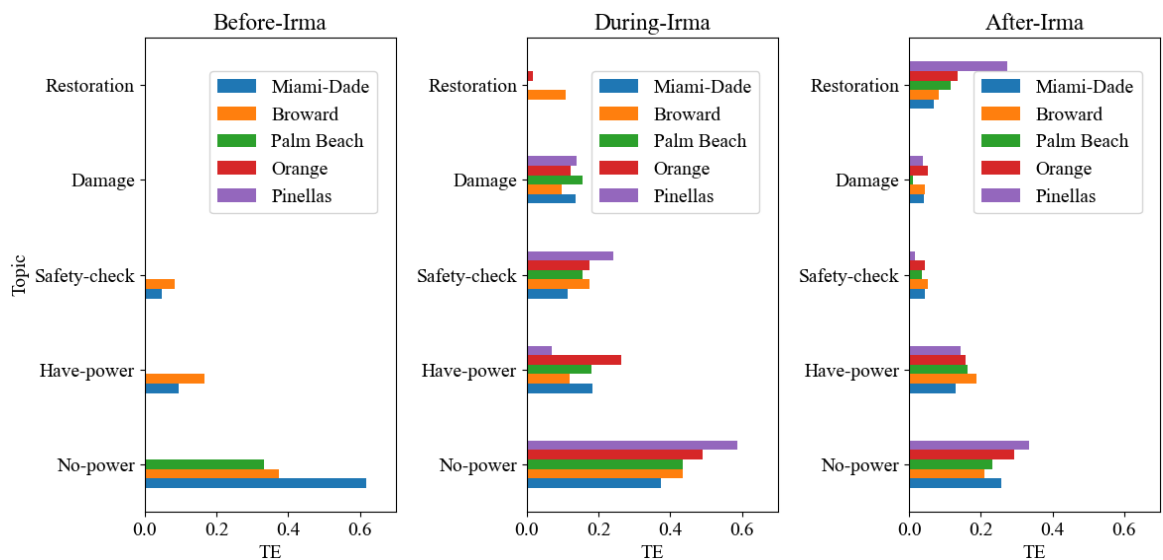

**Figure 7. Spatiotemporal Patterns of the Five Aggregated Topics**

**DISCUSSION**

While extensive studies have focused on using social media to form and enhance situational awareness in disasters, social media-enabled situational assessment of electricity infrastructure has been less investigated. This research examines the effectiveness of social media for assessing electricity infrastructure conditions, which contributes to the domain of infrastructure management in disasters. The results demonstrated that the assessed electricity infrastructure conditions are consistent with the physical facts. In contrast to the traditional approaches (e.g., field inspections, phone- and internet-based reporting, and smart meters), the proposed approach is capable of providing new and supplementary insights on infrastructure conditions as the employed data source is different than that used in traditional approaches. In the scenario wherein traditional approaches are limited due to a variety of reasons (e.g., unreachable communities caused by blocked transportation, overloaded phone-based reporting, and malfunctioned smart meters), the proposed approach can be used to provide timely insights on infrastructure conditions. While traditional approaches are functioning well, the proposed approach can be used to cross-validate those indications from traditional approaches or to provide supplemental insights. Such capabilities of the proposed approach are desired to enhance situational assessment of critical infrastructure in disasters.

**CONCLUSION**

In this research, a systematic and customized approach is developed to sense electricity infrastructure conditions through mining public topics from social media. Electricity infrastructures in Florida impacted by Hurricane Irma are studied for illustration and demonstration. Results show that the developed approach is capable of reliably sensing infrastructure conditions by investigating the spatiotemporal patterns of the modeled public topics.

Academically, this research achieves theoretical and methodological advancements in utilizing social media for sensing electricity infrastructure conditions. From the theoretical side, the research identifies the existing electricity-related public topics in social media, which is beneficial for researchers to further utilize social media data for detailed sensing of electricity infrastructures, such as rapid damage assessment, information update, and restoration process evaluation. From the methodological side, a systematic and customized approach is developed to mine public topics on electricity infrastructure from social media, wherein (1) a fine-tuned BERT-based classification model is developed to identify electricity-related tweets from social media, and (2) public topics on electricity infrastructures are modeled with the most frequent unigrams, bigrams, and trigrams to incorporate the formulaic expressions used by social media users for describing their electricity conditions. Practically, the developed approach can be used to assist infrastructure operators to obtain a timely overview of electricity infrastructure conditions, which contributes to the planning of infrastructure restorations.

Although this research develops a systematic and customized approach for mining public topics on electricity infrastructure from social media, its capability is limited in counties with very few Twitter activities. The main reason for this limitation is that only geotagged tweets, which account for around 1% of the entire tweets (Leetaru, 2019), are employed in this research. In the future, data enrichment methods (profile-based and content-based) will be incorporated to extract more electricity-related social media data, which is beneficial for achieving more reliable modeling of public topics on electricity infrastructures.

**ACKNOWLEDGMENTS**

This study is funded by the Thomas F. and Kate Miller Jeffress Memorial Trust (Grant No. 223476). Any opinions, findings, and conclusions or recommendations expressed in this article are those of the authors and do not necessarily reflect the views of the Thomas F. and Kate Miller Jeffress Memorial Trust.